\def\Bbb#1{{\bf #1}}

\def\fnote#1{\footnote}
\def\blacksquare{\hbox{\vrule width 4pt height 4pt depth 0pt}}

\def\cwleftpar#1#2{\leftskip #1 \rightskip #2 plus 1fill}
\def\cwrightpar#1#2{\leftskip #1 plus 1fill \rightskip #2}
\def\cwcenterpar#1#2{\leftskip #1 plus 1fill \rightskip #2 plus 1fill}
\def\cwfullpar#1#2{\leftskip#1\rightskip#2}

\def\cwoutdent#1#2{\llap{\hbox to #1{#2 \hss}}\ignorespaces}
\def\cwparbegin#1#2#3#4#5{
	\ifcase #1 \cwleftpar{#2}{#3}
	\or \cwrightpar{#2}{#3}
	\or \cwcenterpar{#2}{#3}
	\else \cwfullpar{#2}{#3}\fi
	\ifcase #4 \baselineskip = 1.5\baselineskip
	\or \baselineskip = 2\baselineskip
	\or \baselineskip = 3\baselineskip
	\else \baselineskip = 1\baselineskip\fi
	\ifdim #5 > 0in \else \noindent \fi
	\noindent\ignorespaces}
\documentclass{article}
\begin{document}
\advance \vsize by -1\baselineskip
\def\makefootline{
{\vskip \baselineskip \noindent \folio                                  \par
}}

\vspace{6ex}
\noindent {\Huge Deviation Equations in\\[1ex]
 Spaces with a Transport along Paths}\\[6ex]

\medskip
\noindent Bozhidar Zakhariev Iliev
\fnote{0}{\noindent $^{\hbox{}}$Permanent address:
Laboratory of Mathematical Modeling in Physics,
Institute for Nuclear Research and \mbox{Nuclear} Energy,
Bulgarian Academy of Sciences,
Boul.\ Tzarigradsko chauss\'ee~72, 1784 Sofia, Bulgaria\\
\indent E-mail address: bozho@inrne.bas.bg\\
\indent URL: http://theo.inrne.bas.bg/$^\sim$bozho/}

\vspace{6ex}

\noindent
{\bf Published: Communication JINR, E2-94-40, Dubna, 1994}\\[3ex]

\noindent http://www.arXiv.org e-Print archive No.~math-ph/0303002\\[3ex]

\noindent
{\small
The \LaTeXe\ source file of this paper was produced by converting a
ChiWriter 3.16 source file into
ChiWriter 4.0 file and then converting the latter file into a
\LaTeX\ 2.09 source file, which was manually edited for correcting numerous
errors and for improving the appearance of the text.  As a result of this
procedure, some errors in the text may exist.
}\\[7ex]

\noindent {\bf 1. INTRODUCTION} \par
\medskip
This paper starts a series of applications of the transports along paths
introduced in [1,2] in fibre bundles to certain physical problems.\par All
considerations in the present work are made in a (real) differentiable
manifold $M$ [3,4] whose tangent bundle $(T(M),\pi ,M)$ is endowed with a
transport along paths [1]. Here $T(M):=\cup_{x\in M}T_{x}(M)$,
$T_{x}(M)$ being the tangent to $M$ space at $x\in M$ and $\pi :T(M)
\to M$ is such that $\pi (V):=x$ for $V\in
T_{x}(M)$.\par The set of all sections of a fibre bundle $\xi  [3,4]$ is
denoted by Sec$(\xi )$; e.g.  Sec$(T(M),\pi ,M)$ is the module of vector
fields on M.\par By $J$ and $\gamma :J  \to M$
denoted are, respectively, an arbitrary real interval and a path in M. If
$\gamma $ is of class $C^{1}$, its tangent vector is written as $\dot\gamma$.

The transport along paths in $(T(M),\pi ,M)$ (cf. [1]) is a map
$I:\gamma \mapsto I^{\gamma },
I^{\gamma}:(s,t)\mapsto I^{\gamma }_{s  \to t},
s,t\in J$
being the transport along $\gamma $, where
$I^{\gamma }_{s  \to t}:T_{\gamma (s)}(M)  \to T_{\gamma (t)}(M)$,
satisfy the equalities
\[
^{ }_{ } I^{\gamma }_{t  \to r}\circ I^{\gamma }_{s  \to t}=I^{\gamma }_{s
\to r}, r,s,t\in J,\qquad (1.1)
\]
\[
{ } ^{ }_{ } I^{\gamma }_{s  \to s}={\it id}_{T_{\gamma (s)}}, s\in J.
\qquad(1.2)
\]
Here
{\it id}$_{X}$ is the identity map of the set X.\par A linear transport
$(L$-transport) along paths $L$ in $(T(M),\pi ,M)$ satisfies, besides (1.1)
and (1.2), the equality (cf. [2])
\[^{ }_{ } L^{\gamma }_{s\to t}(u^{i}e_{i}(s))
= H^{i}_{.j}(t,s;\gamma )u^{j}e_{i}(t), s,t\in J, u^{i}\in {\Bbb R}.
\qquad (1.3)
\]
Here and henceforth in our text the Latin
indices run from 1 to $n:=\dim(M)$ and summation from 1 to $n$ is assumed
over repeated indices on different levels; $\{e_{i}(s)\}$ is a basis in
$T_{\gamma (s)}(M)$; and $H(t,s;\gamma ):=  H^{i}_{.j}(t,s;\gamma )  $ is the
matrix of the $L$-transport, in terms of which (1.2) reads\par
\[
^{}_{ } H(s,s;\gamma )={\Bbb I}:=diag(1,\ldots  1):=  \delta ^{i}_{j}
,\qquad (1.2^\prime )
\]
$\delta ^{i}_{j}$being the Kroneker's delta
symbols.\par This work is organized as follows. In Sect. 2,  based  on  the
ideas of $[5,6], a$ strict definition is given  of  the  displacement vector
in a manifold with a transport along paths  in  its  tangent bundle.The
deviation vector between two paths  with  respect  to  a third one is
introduced on the ground of  this  concept.  Sect.  3, which follows the
works $[5,7-10]$, is devoted to the deviation equation, satisfied by the
deviation vector, which is a  generalization of the geodesic deviation
equation (known also in the  mathematical literature as a Jacobi equation).
Special cases  of  this  equation are considered. In particular, it is proved
that it generalizes the equation of motion of two point particles, i.e. the
second  Newton's low of mechanics.

\newpage

\noindent {\bf 2. DISPLACEMENT AND DEVIATION VECTORS}\\

	Let in the tangent fibre bundle $(T(M),\pi ,M)$ to the
differentiable manifold $M$ there be given a transport along paths I and
$\gamma :J  \to M$ be a smooth, of class $C^{1}$, path in M. We define
maps
\[
^{ }_{ } d^{\gamma }_{s}:J  \to T_{\gamma (s)}(M)
=\pi^{-1}(\gamma (s)), s\in J,\qquad (2.1a)
\]
such that
\[
^{ }_{ } d^{\gamma }_{s}(t)
:= \bigl(I^{\gamma }_{u  \to s}\gamma (u)\bigr)du, s,t\in J.
\qquad (2.1b)
\]
{\bf Proposition 2.1.} If I coincides
with some linear transport along paths $L$, then
\[
^{ }_{ } d^{\gamma }_{r}(s)
=d^{\gamma }_{r}(t)+L^{\gamma }_{t  \to r}(d^{\gamma }_{t}(s)),
r,s,t\in J. \qquad (2.2)
\]

{\bf Proof.} (2.2) follows from
(2.1b) and (1.1):\par \noindent $^{ }_{ } d^{\gamma }_{r}(s):=^{s}_{
}\bigl(L^{\gamma }_{u  \to r}$  $(u)\bigr)du
=^{t}_{  }\bigl( L^{\gamma }_{u  \to r}$
$(u)\bigr)du+^{s}_{  }\bigl(L_{t  \to
r}\circ L^{\gamma }_{u  \to t}$  $(u)\bigr)du=$ \par \noindent $^{ }_{ }
=d^{\gamma }_{r}(t)+L^{\gamma }_{t  \to r}(d^{\gamma }_{t}(s)), r,s,t\in $J.
\blacksquare

{\bf Definition 2.1.} The vector $d^{\gamma }_{s}(t)$ will
be called a displacement vector of $\gamma (t)$ with respect to $\gamma (s)$
if $d^{\gamma }_{s}, s\in J$ maps $J$ homeomorphicly onto its image
$d^{\gamma }_{s}(J)$.\par Generally, for an arbitrary transport along paths
and a path $\gamma $ the condition in this definition is not fulfilled. But
it happens that under sufficiently general conditions there exist suitable
combinations of I and $\gamma $ for which it is true. Without going into
details of this problem, we shall present only two examples for such cases.
They are expressed by the proved below proposition 2.3 and corollary 2.1 and
practically include all combinations essential for physical application. For
the first example we need

{\bf Definition 2.2.} Let I be a transport
along paths in $(T(M),\pi ,M)$. An I-path is a smooth, of class $C^{1}$, path
$\gamma :J  \to M$ the tangent vector field
$\dot\gamma\in Sec(T(\gamma (J)),\pi ,\gamma (J))$
of which is transported by means of I along $\gamma $,
i.e.
\[
\dot\gamma(t)=I^{\gamma }_{s  \to t}  \dot\gamma(s), s,t\in J. \qquad  (2.3)
\]
The existence problem for the I-paths in the case of $L$-transports along
paths, i.e. for $L$-paths, is shortly formulated in Ref. [11].\par As the
theory of I-paths is not in the main direction of this investigation, we
shall only remark that an evident special case of the I-paths (in affine
parameterization) in manifolds with connection is the geodesic paths, whose
tangent vector undergoes a parallel transport (defined by the manifold's
connection) along themselves [3,4].

{\bf Proposition 2.2.} If $\gamma :J  \to M$ is an I-path, then
\[
d^{\gamma }_{s}(t)=(t-s)  \dot\gamma(s), s,t\in J. \qquad (2.4a)
\]
{\bf Proof.} (2.4a) follows from the substitution of (2.3) for $s=u$ and
$t=s$ into (2.1b).\blacksquare

{\bf Corollary 2.1.} If $\gamma :J  \to M$ is a regular I-path, then $d^{\gamma }_{s}(t)$ is a displacement vector of $\gamma (t)$ with respect to $\gamma (s)$, i.e. the condition in definition 2.1 is fulfilled.\par

{\bf Proof.} From (2.4a) it follows that for any $s\in J$ the mapping (2.1a)
is linear, so, due to the regularity of $\gamma  ($i.e.   $(s)\neq 0)$, it is
a diffeomorphism, and consequently homeomorphism, from $J$ onto $d^{\gamma
}_{s}(J).\blacksquare $

{\bf Proposition 2.3.} If $\gamma $ is a $C^{1}$ path without
self-intersections, then in the case of $L$-transports along paths the
mappings $d^{\gamma }_{s}, s\in J$ map $J$ locally homeomorphicly on its
image $d^{\gamma }_{s}(J)$, i.e. locally $d^{\gamma }_{s}(t)$ is a
displacement vector of $\gamma (t)$ with respect to $\gamma (s)$.

{\bf Remark.} In this case the word "locally" means in some part of
(or over the whole) set $\gamma (J)$ in a neighborhood of which there exist
local coordinates with the properties described in [12], lemma 7. (See also
below the proof of this proposition.)

{\bf Proof.} Firstly we shall prove
that $d^{\gamma }_{s}, s\in J$ are locally injective, i.e. if $t_{1}\neq
t_{2}$, then $d^{\gamma }_{s}(t_{1})\neq d^{\gamma }_{s}(t_{2})$. In fact,
for linear transports along paths, by proposition 3.1 of [2] in $(T(M),\pi
,M)$ along $\gamma $ there exists a basis $\{E_{i^\prime }\}$, which by [12],
lemma 7 is (locally) holonomic and in which the matrix of the transport is $
H^{i^\prime }_{..j^\prime }(t,s;\gamma )  =  \delta ^{i}_{j}  $. In
this basis,  if $\gamma (s)$ and $\gamma (t)$ belong to one and the same
coordinate neighborhood, we have
\[
(d^{\gamma }_{s}(t))^{i^\prime }
=\int^{t}_{s}H^{i^\prime }_{..j^\prime }(s,u;\gamma )
\dot\gamma^{j^\prime }(u)du
=\int^{t}_{s} \dot\gamma^{i^\prime }(u)du
=\gamma ^{i^\prime }(t)-\gamma ^{i^\prime }(s),
\qquad (2.4b)
\]
where the validity of the last equality
follows from the fact that $\gamma $ is without self-intersections.
Consequently, if $\gamma (s), \gamma (t_{1})$ and $\gamma (t_{2})$ belong to
one and the same coordinate neighborhood, then $d^{\gamma }_{s}(t_{1})\neq
d^{\gamma }_{s}(t_{2})$ is equivalent to $\gamma ^{i^\prime }(t_{1})\neq
\gamma ^{i^\prime }(t_{2}), t_{1},t_{2}\in J$, which is equivalent to
$t_{1}\neq t_{2}$ only if the path $\gamma $ is without selfintersection in
the mentioned coordinate neighborhood, as is supposed here.\par The maps
$d^{\gamma }_{s}$, evidently, are locally (in the above neighborhood) unique
and differentiable, besides, due to (2.1b), we have
$\frac{d}{d t}(d^{\gamma }_{s}(t))=I^{\gamma }_{t  \to s}\dot\gamma(t)$. The
existence and the continuity of $(d^{\gamma }_{s})^{-1}:d^{\gamma }_{s}(J)
\to J$ follows from the representation (2.4b) of $d^{\gamma }_{s}(t)$ in the
basis $\{E_{i^\prime }\}$.

Analogously the proposition can be proved when
only the points $\gamma (t_{1})$ and $\gamma (t_{2})$ lie in the same
coordinate neighborhood with the needed properties. The only difference now
is that if $\gamma (s)$ is out of this neighborhood, then in the
right-hand-side of the last equality in (2.4b) there appear terms
independent of $t$, which does not change the validity of the above
considerations.\blacksquare

{\bf Remark.} If the path $\gamma $ has self-intersections, in the
right-hand-side of (2.4b) the term
$\sum_k\oint_{\gamma_k}\dot\gamma^{i^\prime }(u)du$
appears, where the summation is taken over all closed loops
$\gamma_{k}$ formed by the restriction of $\gamma $ on the interval
$[\min(s,t),\max(s,t)]$. Therefore in the general case, from (2.4b) it does
not follow that $d^{\gamma }_{s}$are injective maps.

Further, the transport I and the path $\gamma $ are supposed to be chosen so that the condition in definition 2.1 would be true, i.e. $d^{\gamma }_{s}(t)$ would be a displacement vector of $\gamma (t)$ with respect to $\gamma (s)$.

The displacement vector, a direct generalization of the difference of two
Euclidean radius-vectors (see below Sect. 4), finds application due to the
property that it has the meaning of "vector relative coordinate" on the
one-dimensional submanifold $\gamma (J)$, i.e. if a point $\gamma (s)\in
\gamma (J)$ for fixed $s\in J$ is given, then from the knowledge of the
displacement vector $d^{\gamma }_{s}(t)$ for any $t\in J$ one can define
(recover) the point $\gamma (t)$ and vice versa. In fact, if $t\in J$, then
by (2.1b) to it there corresponds a unique vector $d^{\gamma }_{s}(t)\in
T_{\gamma (s)}(M)$ and on the opposite, if $\Delta \in d^{\gamma }_{s}(J)$,
then as $d^{\gamma }_{s}:J  \to d^{\gamma }_{s}(J)$ are homeomorphisms there
exists a unique $t\in J$, and so a point $\gamma (t)\in d^{\gamma }_{s}(J)$,
with the property $d^{\gamma }_{s}(t):=\Delta $. For the same reason, with
the help of a displacement vector there can be defined also a (global) chart
on $\gamma (J)$: because $d^{\gamma }_{s}:J  \to d^{\gamma }_{s}(J)$ is a
homeomorphism, the set $d^{\gamma }_{s}(J)\subset T_{\gamma (s)}(M)$ is one
dimensional submanifold and, hence, there exists a homeomorphism $\varphi
_{s}:d^{\gamma }_{s}(J)  \to {\Bbb R}^{1}$, as a consequence of which
$(\gamma (J),\varphi _{s}\circ d^{\gamma }_{s}\circ \gamma ^{-1})$, where
$\gamma ^{-1}:  \gamma (J)  \to J$ and $\gamma ^{-1}(\gamma (t)):=t, t\in J$,
is a global chart on $\gamma (J)$.

Using the displacement vector one can construct the so called deviation
vector between two paths with respect to a third one. This is done as
follows.

Let there be given paths $x_{a}:J_{a}  \to M, a=1,2$ and $x:J  \to $M. Let
there be fixed one-to-one maps $\tau _{a}:J  \to J_{a}, a=1,2$. (These maps
always exit as all real intervals are equipollent.) Let also be given the one
parameter families of paths $\{\gamma _{s}: \gamma _{s}:J$  $  \to M, s\in
J\}$ and $\{\eta _{s}: \eta _{s}:J$  $  \to M, s\in J\}$ having the
properties $\gamma _{s}(r$  $):=x_{1}(\tau _{1}(s)):=\eta _{s}(t$  $), \gamma
_{s}(r$  $):=x_{2}(\tau _{2}(s))$ and $\eta _{s}(t$  $):=x(s)$ for some $r$
,$r$  $\in J$   and $t$  ,$t\in J$  , $s\in J$. The paths $\gamma _{s},
s\in J$ are supposed smooth and such that the maps $d^{\gamma _{s}}_{r}, r\in
J$  , $s\in J$ determined by them from (2.1) define corresponding
displacement vectors.

{\bf Definition 2.3.} The deviation vector of $x_{2}$ with respect to $x_{1}$
relatively to $x$ at the point $x(s), s\in J$ is the vector

\noindent
$h_{21}
:=h_{21}(s;x)
:=\left(\begin{array}{c} \end{array}\right.
	I^{\eta _{s}}_{t^\prime_{s}   \to t^{\prime\prime}_{s}}\circ
d^{\gamma _{s  }}_{r^\prime _{s}} ) (r^{\prime\prime}_s  )=$
\par
$^{ }_{ } =I^{\eta _{s}}_{t^\prime_{s}   \to t^{\prime\prime}_{s}}$
$\int^{r^{\prime\prime}_s}_{r_s^\prime }
\bigl(I^{\gamma _{s}}_{u  \to r^\prime _{s}}$
$\dot\gamma_{s}(u)\bigr)du \in T_{x(s)}(M)\qquad (2.5)$

The deviation
vector and the objects involved in its definition can be interpreted from the
view point of the physical applications as follows. (Anything written below
needs many additional definitions and precise statements as to have a strict
meaning. For this reason one may think that $M$ in it is the 4-dimensional
space-time $V_{4}$ of general relativity - see e.g. [13].) We can interpret
the paths $x_{1}$ and $x_{2}$ as trajectories (world lines) of two observed
point particles, the path $x$ - as a trajectory of an observer "studying"
their behavior. The parameters $s_{1}\in J_{1}, s_{2}\in J_{2}$and $s\in J$
may be considered as "proper times" of the corresponding particles. The maps
$\tau _{1}$ and $\tau _{2} $give the connection between these proper times,
define the "observation process" in this concrete situation, and, in a
certain sense, they give some "simultaneity" between all particles:
$\tau_{1}$ and $\tau _{2}$ define a simultaneity between the observer and the
observed particles and $\tau _{2}\circ \tau ^{-1}_{1}$- between the observed
particles. For a fixed $s\in J$ the paths $\gamma _{s}$ and $\eta _{s}$ can be
regarded as trajectories (world lines) of "signals" which "physically
realize" the maps $\tau _{2}\circ \tau ^{-1}_{1}$ and $\tau _{1}. ($ For
instance, in $V_{4}$if $\gamma _{s}$and $\eta _{s}$ are isotropic geodesic
paths, then the above described construction corresponds to the definition of
simultaneity with the help of light signals - see [13].) In this context the
deviation vector the meaning of a vector describing the relative position of
the second observed particle with respect to the first one as this is "seen"
from an observer.

At the end of this section we want to present the lowest and,
respectively, the most used approximations when one works with the
displacement and deviation vectors.

If the transport I has a continues dependence on (one of) its parameters,
then using the formula $\int^{b}_{ a}f(u)du=f(a)(b-a)+O((b-a)^{2})$ for
any continues function $f:[a,b]  \to {\Bbb R}$, from (2.1b) and (1.2), we
find
\[
d^{\gamma }_{s}(t)= (t-s)\dot\gamma(s) +O((t-s)^{2}).\qquad (2.6)
\]
If the  points $\gamma (s)$ and $\gamma (t)$ are "sufficiently"
(infinitesimally) close, then the vector
\[
\zeta ^{\gamma }_{s}(t):=(t-s)\dot\gamma(s)\qquad (2.7)
\]
is a "good" (of first order with respect to $t-s)$ approximation to the
displacement vector (2.1b).  By definition it is called the {\it
infinitesimal displacement vector}.  Evidently, in the case of I-paths, due
to proposition 2.2, the vector (2.7) coincides (globally) with the
displacement vector.

From (2.5) and (2.6), we find the following representation of the deviation
vector
\[
h_{21}=I^{\eta _{s}}_{t^\prime_{s} \to t^{\prime\prime}_{s}}
[\dot\gamma_{s}(r_s^\prime )(r_s^{\prime\prime} -r_s^{\prime})
+O((r_s^{\prime\prime} -r_s^{\prime})^{2})],
\qquad (2.8)
\]
which for $L$-transports in local coordinates, as a consequence of (1.2), is
equivalent to
\[
h^{i}_{21}
= \dot\gamma^{i}_{s}(r'_s)
(r_s^{\prime\prime} -r_s^{\prime})+O(t_s^{\prime\prime} -t_s^{\prime})
+O(( r_s^{\prime\prime} -r_s^{\prime} )^{2}).
\qquad (2.9)
\]
Here we see that within the quantities of first order with respect to
$(t_s^{\prime\prime} -t_s^{\prime} )$ and second order with respect to
$(r_s^{\prime\prime} -r_s^{\prime} )$ the vector
\[
\zeta _{21}:=(r_s^{\prime\prime} -r_s^{\prime})  \dot\gamma_{s}(r'_s ),
\qquad (2.10)
\]
which, though being defined at another point, by its
components is an approximation to the deviation vector (2.5). In this case
the vector (2.10) is called the {\it infinitesimal deviation vector} [13].

\par
\medskip
\medskip
\noindent {\bf 3. DEVIATION EQUATIONS} \par
\medskip
Let the manifold $M$ be endowed with a linear connection (covariant
derivative) $\nabla $ with local coefficients $\{\Gamma ^{i}_{.jk}(x)\}$
$(cf.  [3,4])$.  If $X,Y,Z\in $Sec$(T(M),\pi ,M)$, then the tensors
(operators) of torsion $T$ and curvature $R$ are
\[
 T(X,Y):=\nabla _{X}Y-\nabla _{Y}X-[X,Y],\qquad (3.1)
\]
\[
 R(X,Y)Z:=\nabla _{X}\nabla _{Y}Z-\nabla _{Y}\nabla
_{X}Z-\nabla _{[X,Y]}Z,	\qquad (3.2)
\]
where $[X,Y]$ is the commutator of $X$ and $Y$,
and in any local basis $\{E_{i}\}$ we have
$[E_{i},E_{j}]=:C^{k}_{.ij}E_{k}$, so their components, respectively, are:
\[
T^{i}_{.jk}=-2\Gamma ^{i}_{.[jk]}-C^{i}_{.jk},\qquad (3.3)
\]
\[
R^{i}_{\hbox{.jkl}}=-2\Gamma ^{i}_{.j[k,l]}-2\Gamma
^{m}_{.j[k}\Gamma ^{i}_{.  m  l]}-\Gamma ^{i}_{jm}C^{m}_{.kl}.\qquad (3.4)
\]
Let us express $\nabla _{x}Z$ from equation (3.1) for $Y=Z$, substitute the
obtained result into (3.2), and put in the thus found equality $X=\xi ,
Y=Z=U$ for $\xi ,U\in $Sec$(T(M),\pi ,M)$. Thus using the skewsymmetry of $T,
R$ and the commutator on their first two arguments, we get the equality \[
\nabla ^{2}_{U}\xi =R(U,\xi )U+\nabla _{\xi }(\nabla _{U}U)+\nabla
_{U}(T(U,\xi ))+\nabla _{U}[U,\xi ]+\nabla _{[U,\xi ]}U.
\qquad (3.5)
\]
In [14] this equality is called the $^{\prime\prime}$basic
equation$^{\prime\prime}$ as from it by imposing additional condition on the
quantities involved in it the deviation equations used in the literature can
be obtained (for geodesic as well as for nongeodesic paths) [14,7,8].

The physical meaning and interpretation of the equality (3.5) can be
obtained as follows.

Let besides the connection in $(T(M),\pi ,M)$ there be defined a transport
along paths I and there be given the construction of paths,
${\Bbb R}$-intervals and maps between them appearing in definition 2.3 of the
deviation vector (2.5), for which we suppose to have $a C^{2}$ dependence on
$s\in $J.

Let us put $U$ in (3.5) to be the tangent vector field to the path $x$ and
$\xi $ to be the field of the deviation vector of $x_{2}$with respect to
$x_{1}$, i.e.
\[
U_{x(s)}=\dot{x}(s), \quad  \xi _{x(s)}=h_{21}(s;x).\qquad (3.6)
\]
Then on $x(J)$
\[
\nabla _{U}= \frac{D}{ds} \Big|_{x}\qquad (3.7)
\]
is the covariant differentiation along $x$ and (3.5) takes the form
\[
\frac{D^2}{ds^2}\Big|_{x }h_{21}=R(U,h_{21})U+\nabla _{h_{21}}\bigl(
\frac{D}{ds}\Big|_{x}U\bigr)+
\frac{D}{ds} \Big|_{x}\bigl(T(U,h_{21})\bigr)+
\]
\[
\qquad + \frac{D}{ds}\Big|_{x}[U,h_{21}]+\nabla _{[U,h_{21}}U.
\qquad (3.8)
\]
The equality (3.8) is called the {\it generalized deviation equation}.
In the local case, i.e. when $h_{21} $ is an infinitesimal vector, which
usually is identified with the infinitesimal deviation vector (2.10), this
name was introduced in [15,8,14], and in the global case, i.e. for an
arbitrary deviation vector $h_{21}$, in [5].

The physical interpretation of the generalized deviation equation (3.8) may
be found, for example, in [5,8,9] and it is based on the physical
interpretation of the deviation vector given in Sect. 2. Due to it $h_{21},
\nabla _{U}h_{21}$and $\nabla ^{2}_{U}h_{21}$ are interpreted, respectively,
as relative coordinate, velocity and acceleration (or, more precisely, these
are the deviation vector, the deviation velocity and the deviation
acceleration, but for the moment this is not essential) of the second
observed particle with respect to the first one relatively to the observer.
The quantities $U$ and $\nabla _{U}U$ are interpreted, respectively, as the
velocity and force per unit mass acting on the observer. As a consequence of
this we can say that the generalized deviation equation (3.8) gives the
relative acceleration $\nabla ^{2}_{U}h_{21}$between the observed particles
as a function of the characteristics of the manifold $M (R, T$ and $\Gamma
^{i}_{.jk})$, the trajectory (the world line) of an observer $(s, x, U$ and
$\nabla _{U}U)$ and the relative movement of the observed particles
$(h_{21}$and $\nabla _{U}h_{21})$.

{\bf Example 3.1.} Now, analogously to the investigations in [9], on the
basis of the above general considerations, we shall derive the nonlocal
(noninfinitesimal) deviation equation of the geodesics.

Let $y:\Lambda _{2}  \to M$, where $\Lambda _{2}$is a neighborhood in ${\Bbb
R}^{2}$, be $a C^{2}$ congruence of geodesics (with respect to the connection
of $M)$ paths. This means that the tangent vectors $U$ and $V$, respectively,
to the $u$-paths $y(\cdot ,v), v=$const and $v$-paths $y(u,\cdot ), u=$const,
$(u,v)\in \Lambda _{2}$which are geodesics, satisfy the equalities
\[
\nabla _{U}U_{(u,v)}=f_{v}(u)U_{(u,v)}, \quad
\nabla_{V}V_{(u,v)}=g_{u}(v)V_{(u,v)}.\qquad (3.9)
\]
 ere the restriction
$|_{(u,v)}$means that the corresponding quantities are taken at the point
$y(u,v), (u,v)\in \Lambda _{2}$ and the functions $f_{v}$and $g_{u}$depend
only on the choice of the parameters $u$ and $v$ (for instance, if $u$ is an
affine parameter, then by definition $f_{v}(u)\equiv 0)$.

We have to find the deviation equation of two arbitrary $u$-paths from the
family $y(\cdot ,v), v=$const, $(u,v)\in \Lambda _{2}$. For this purpose, in
the above general construction, we substitute: $y(\cdot ,v_{1})$ and $y(\cdot
,v_{2})$ for some fixed values $v_{1}$and $v_{2}$of the parameter $v$,
respectively, for $x_{1}$and $x_{2}; y(\cdot ,v_{1})$ for $x ($and
consequently $\tau _{1}=\tau _{2}=${\it id}); and $y(u,\cdot )$ for $\gamma
_{s}$. As a concrete and "most natural" realization of the transport I we
shall use the parallel transport defined by the connection of $M$ will be
used.

As $y(u,\cdot )$ is a geodesic, we have
\[
I^{y(u,\cdot )}_{v_{0}}(V_{(u,v_{0}})=\mu _{u}(v_{0},v)V_{(u,v)}
\]
for some scalar function $\mu $ of $u, v_{0}$and $v$, which due to (1.2) has
the property $\mu _{u}(v_{0},v_{0})=1$. On the other hand, $(cf. [2]$,
proposition 4.1), the fact that $I^{y(u,\cdot )}$is a parallel transport
along $y(u,\cdot )$ leads to $\nabla _{V}$  $_{(u,v)}\circ I^{y(u,\cdot
)}_{v_{0}}\equiv 0$. Combining these equalities with the second equation from
(3.9), we get $\mu _{u}(v_{0},v)=\exp\bigl(-^{  ^{v}}_{
v_{0}}g_{u}(w)dw\bigr)$. Due to this from (2.5), we find the deviation
vector of $y(\cdot ,v_{2})$ with respect to $y(\cdot ,v_{1})$ at the point
$y(u,v_{1})$ as
\[
h:=h(u,v_{1},v_{2}):=\lambda \cdot V_{(u,v_{1})},\qquad (3.10)
\]
\[
\lambda
:=\lambda_{u}(v_{1},v_{2})
:=\int^{v_2}_{v_1}\exp{\bigl(}- \int_{v_1}^{v}g_{u}(w)dw\bigr) dv
=
\frac{a_u(v_2)-_u(v_1)}{\partial a_u(v_1)/\partial v_1}
\qquad (3.11)
\]
where
$a_{u}(v):=C_{1}(u) \int^{v}_{v_0}
\exp\bigl(-\int^{t}_{v_0}g_{u}(w)dw\bigr) dt+C_{2}(u)$, with
$C_{1}\neq 0$, and $C_{2}$being arbitrary functions, is any affine parameter
of $y(u,\cdot )$.

The form of the deviation equation (3.8) in the considered case is defined
by two additional conditions. First, on $y(\cdot ,v_{1})$ the first equation
of (3.9) gives
\[
\nabla _{U}U_{(u,v_{1}}=f_{v_{1}}(u)U_{(u,v_{1})}\qquad (3.12)
\]
Second, as $u$ and $v$ are independent parameters of the $C^{2}$congruence
$y$, in local coordinates, we get $\partial ^{2}y^{i}(u,v)/\partial u\partial
v= =\partial ^{2}y^{i}(u,v)/\partial v\partial u$, which on $y(\cdot ,v_{1})$
reduces to
\[
[h,U]_{(u,v_{1}}
=L_{h}U_{(u,v_{1}}=-\lambda ^\prime V_{(u,v_{1})}
=-h\lambda ^\prime/\lambda ,\qquad (3.13)
\]
where $\lambda ^\prime :=\partial \lambda /\partial
u$ and $L_{h}U:=[h,U]$ is the commutator of $h$ and $U ($or the Lie
derivative of $U$ with respect to $h)$. Substituting (3.12) and (3.13) into
(3.8) and using the notation $\lambda ^{\prime\prime}:=\partial \lambda
^\prime /\partial u$ and the relationship $L_{h}U=\nabla _{h}U-\nabla
_{U}h-T(h,U)$, which is true for any vector fields $h$ and $U ($see (3.1)),
we find the geodesic deviation equation as
\[
\frac{D^2}{du^2}\Big|_{ y(\cdot ,v_{1})} h_{(u,v_{1}}
=R(U,h) U_{(u,v_{1})}+
\Bigl[ f_{v_{1}}(u)T(U,h)+(\nabla _{U}T)(U,h)+
\]
\[
+T(U, \frac{D}{du}\Big|_{y(\cdot ,v_{1})}h) \Bigr]_{(u,v_{1})}
+\lambda \cdot \Bigl(
\frac{D}{du}\Big|_{y(u,\cdot )}(f_{v}(u)U) \Bigr)_{(u,v_{1})}+
\]
\[
+
\frac{\lambda'}{\lambda}
\Bigl[
2\frac{D}{du}\Big|_{y(\cdot ,v_{1})}h-2 \frac{\lambda'}{\lambda} \cdot h
+T(h,U)\Bigr]\Big|_{(u,v_{1})}
+ \frac{\lambda^{\prime\prime}}{\lambda} \cdot h|_{(u,v_{1})}.
\qquad (3.14)
\]
If the parameter $v$ is affine, then (by definition) $g_{u}(v)=0$, so
now $(3.10), (3.11)$ and (3.14) take, respectively, the form:
\[
h=(v_{2}-v_{1})V_{(u,v_{1})},\quad  \lambda =v_{2}-v_{1},\qquad (3.15)
\]
\[
\frac{D^2}{du^2}\Big|_{y(\cdot ,v_{1})} h_{(u,v_{1})}
=R(U,h)U_{(u,v_{1})}+
\Bigl[
f_{v_{1}}(u)T(U,h)+(\nabla _{U}T)(U,h)+
\]
\[
+T(U, \frac{D}{du}\Big|_{y(\cdot ,v_{1})}h) \Bigr] \Big|_{(u,v_{1})}+
(v_{2}-v_{1})\cdot
\Bigl(\frac{D}{du}\Big|_{y(u,\cdot )}(f_{v}(u)U)
\Bigr) \Big|_{(u,v_{1})}.
\qquad (3.16)
\]
If, besides, $u$ is affine too, then (by definition)
$f_{v}(u)=0$ and (3.16) reduces to the equation
\[
\frac{D^2}{du^2}\Big|_{y(\cdot ,v_{1})}h_{(u,v_{1})}
=R(U,h)U_{(u,v_{1})}+(\nabla _{U}T)(U,h)_{(u,v_{1})}+
\]
\[
\qquad +T(U, \frac{D}{du}\Big|_{y(\cdot ,v_{1}}h)|_{(u,v)}.\qquad (3.17)
\]

	Analogously one can get the deviation equation for the congruence
$y:\Lambda _{2}  \to M$ in the case when only the $v$-paths $y(u,\cdot )$ are
geodesics.  Then, as there remains only the additional condition (3.13) the
deviation vector is also given by $(3.10)-(3.11)$.  So
\[
\frac{D^2}{du^2}\Big|_{y(\cdot ,v_{1})}h_{(u,v_{1})}
= R(U,h)U_{(u,v_{1})}+ \frac{D}{du}\Big|_{y(\cdot ,v_{1}}[T(U,h)|_{(u,v_{1})}+
\]
\[
\Bigl(
\frac{D}{du}\Big|_{y(\cdot ,v)} \frac{D}{du}\Big|_{y(u,\cdot )}U
\Bigr)
\Big|_{(u,v_{1}}+ \frac{\lambda'}{\lambda}\cdot
[2 \frac{D}{du}\Big|_{y(\cdot ,v_{1})}h - 2 \frac{\lambda'}{\lambda}\cdot h +
\]
\[
+ T(h,U)]_{(u,v_{1})}+ \frac{\lambda^{\prime\prime}}{\lambda} \cdot
h_{(u,v_{1})},
\qquad (3.18)
\]
If we impose also the condition (3.12), we see
that (3.18) reduces to (3.14).

{\bf Example 3.2.} In this example, based on the work [10], we shall show
that the deviation equation (3.8) contains as its special case the equation
of relative motion of two point particles. In this sense the deviation
equation is a generalization of the second Newton's law of the dynamics.

Let the $C^{2}$ trajectory $x:J  \to M$ of the observer be given as a
solution of the following initial-value problem:
\[
\nabla _{U}U_{x(s)}=F(s,x,U)\in T_{x(s)}(M), \qquad U:= \dot{x},\
s\in J,\qquad (3.19a)
\]
\[
x(s_{0})=x_{0}\in M, \quad U_{x_{0}}:=\dot{x}(s_{0})=U_{0}\in T_{x(s_{0})}(M),
\ s_{0}\in J,\qquad (3.19b)
\]
where $x_{0}$ and $U_{0}$ are fixed, and $F$ is a continuous function of its
arguments.  Physically this means to consider a (point) observer who passes
through the point $x_{0}$ with velocity $U_{0}$ and undergoes a force per
unit mass F.

Let the family of $C^{2}$ paths $\{\gamma _{s}: s\in J\}$ be given as the
unique solution of the following initial-value problem:
\[
\nabla _{\gamma ^\prime }\gamma ^\prime|_{\gamma_{s}(r)}=F_{s}(r)
:=F_{s}(r,\gamma _{s}(r)),\quad \gamma (r)\in T_{\gamma _{s}}(M), \ s\in J,
\qquad (3.20a)
\]
\[
\gamma _{s_{0}}(r)=\chi (r)\in M,\quad \gamma_{s_0}(r)
=\varphi (r)\in T_{\chi(r)}(M),\ s_{0}\in J,
\qquad (3.20b)
\]
where $\gamma ^\prime $ is the tangent vector field to the $s$-paths
$\gamma_{.}(r), r=$const$\in J'_s$, $s\in J$ (i.e.
$(\gamma ^\prime|_{\gamma _{s}(r)})^{i}:=\partial \gamma ^{i}_{s}(r)/\partial
s)$, and $F_{s}, \chi $ and $\varphi $ are continuous functions of their
arguments. Physically $F_{s}$ is interpreted as a force field (force per unit
mass) acting in the two-dimensional region $\{\gamma _{s}(r), r\in J_{s},
s\in J\}$.

	Let us remind (see Sect. 2) that by definition
$\gamma _{s}(r'_s):=x_{1}(\tau _{1}(s)):=\eta _{s}(t)$,
$\gamma_{s}(r_s^{\prime\prime}):=x_{2}(\tau _{2}(s))$ and
$\eta_{s}(t_s^{\prime\prime}):=x(s)$ for some
$r'_s,r_s^{\prime\prime} \in J$, $t',t^{\prime\prime}\in J$,
and $s\in J$.

Further in this example we suppose that the transport I is linear, i.e. we
shall work with $L$-transports (see [2]).

The following purpose is to write, in the considered case, the deviation
equation for the deviation vector $h_{21}$ of $x_{2}$ with respect to
$x_{1}$ relatively to $x$ in the form of equation of motion  that is "most
close" to the second law of the Newton's mechanics. It "more clearly" shows
the dependence of the relative (deviation) acceleration between the observed
particles on the force fields $F$ and $F_{s}$. (This intention comes from the
above given physical interpretation of the deviation equation.)

To write certain formulae compactly, we shall generalize the operation of
differentiation of vector fields along paths (see e.g. (3.7)). Let $p,q\ge 0$
be integers, $z_{a}:J  \to M, a=1,\ldots  ,p+q$ be $C^{1}$paths, $z:J  \to M
\cdot \cdot \cdot   M$ ($p+q$ times) with $z(s):=(z_{1}(s),\ldots
,z_{p+q}(s)), s\in J$ and $T^{p}_{.q}(z(s);M):=T_{z_{1}}(M)\otimes \cdot
\cdot \cdot \otimes T_{z_{p}}(M)\otimes T^{*}_{z_{p+1}}(M)\otimes \cdot \cdot
\cdot \otimes T_{z_{p+q}}(M)$.

For every $s\in J$, we define the map
\[
\frac{D}{ds} : Sec\bigl( \bigcup_{t\in_{J}} T^{p}_{.q}(z(t);M),\pi ,z(J))
\to T^{p}_{.q}(z(s);M),
\]
where $\pi (A_{z(s)}):=z(s)$ for $A_{z(s)}\in T^{p}_{.q}(z(s);M)$, such that
for any section
$A\in $Sec$\bigl(\bigcup_{t\in J}T^{p}_{.q}(z(t);M),\pi,z(J))$,
in local coordinates we have
\[
\Bigl( \frac{D}{ds}A \Bigr)^{i_1\cdots i_p}_{j_1\cdots j_q}
:=
\frac{d}{ds}\Bigl( A^{i_1\cdots i_p}_{j_1\cdots j_q}(z(s) \Bigr) +
\]
\[
+ \sum_{a=1}^{p}\Gamma
^{i_{a}}_{..kl}(z_{a}(s))
A^{i_1\cdots i_{a-1}ki_{a+1}\cdots i_p}_{j_1\cdots j_q}(z(s))
					\dot{z}_{a}^{l}(s) -
\]
\[
- \sum_{b=1}^{q}\Gamma
^{k}_{.j_{b}}(z_{p+b}(s))
A^{i_1\cdots i_p}_{j_1\cdots j_{b-1}ki_{b+1}\cdots j_q}(z(s))
						\dot{z}^{l}_{p+b}(s),
\qquad (3.21)
\]
where $\dot{z}_{a}$ is the tangent vector field to $z_{a}$, $a=1,\ldots ,p+q$.

	With the help of (3.21) it is easy to check that $D/ds$ is a
derivation of the (many-point) tensor algebra over $z(J)$, i.e. this operator
is linear, commutes with the contraction operator (defined now only on
indices referring to dual spaces) and satisfies the relation $D/ds(A\otimes
B)=(DA/ds)\otimes B+A\otimes (DB/ds)$.

	If $p+q=1$, then from (3.21) follows $D/ds=d/ds\mid _{z_{1}}$, i.e.
when acting on vector fields or 1-forms defined over $z_{1}(J)$ the above
defined operator reduces to a covariant differentiation along $z_{1}$.

	Let the $L$-transport along $\gamma:J\to M$ from $s$ to $t$,
$s,t\in J$ in $(T(M),\pi ,M)$ be defined by the matrix $
H^{i}_{.j}(t,s:\gamma )  $ through (1.3) and $\{E_{i}\mid _{y}\}$ be a basis
in $T_{y}(M), y\in $M. We put
\[
H:=H^{i}_{.j}(t_s^{\prime\prime},t_s^{\prime} :\eta_{s})E_{i}
				|_{\eta_{s}(t_s^{\prime\prime}) }
\otimes E^{j}|_{\eta_{s}(t_s^{\prime})}\in
T_{\eta_{s}(t_s^{\prime\prime})}(M)\otimes
T^{*}_{\eta_{s}(t_s^{\prime})}(M),
\]
\[
H^{-1}
:=H^{i}_{.j}(t_s^{\prime} ,t_s^{\prime\prime}:\eta _{s})
E_{i}|_{\eta_{s}(t_s^{\prime})}\otimes
E^{j}|_{\eta_{s}(t_s^{\prime\prime})}\in
T_{\eta_{s}(t_s^{\prime})}(M)\otimes
T^{*}_{\eta_{s}(t_s^{\prime\prime})}(M) ,
\]
\[
\Lambda (r)
:=H^{i}_{.j}(r'_s,r:\gamma _{s})
E_{i}|_{\gamma_{s}(r_s')}\otimes
E^{j}|_{\gamma_{s}(r)}\in
T_{\gamma_{s}(r')}(M)\otimes T^{*}_{\gamma_{s}(r)}(M) .
\]

For brevity, with a point $(\cdot )$ the contracted tensor product will be
denoted, i.e. if $X\in T^{p}_{.q}|_{y}(M)$,
$Y\in T^{p^\prime }_{..q^\prime }|_{y}(M), p^\prime ,q\ge 0$ and $p,q^\prime
\ge 1$, then $X\cdot Y:=C^{p}_{q+1}(X\otimes Y), C^{p}_{q}$ being the
contraction operator on the $p$-th super- and $q$-th subscript.

Using the above notions, we can write the deviation vector (2.5) as
\[
h=
H\cdot \int^{r_s^{\prime\prime}}_{r_s^\prime }\Lambda (u)
\cdot \dot\gamma_{s}(u)du.
\qquad (3.22)
\]
Hence, we find:
\[
\Bigl(\frac{D}{ds}\Big|_x\Bigr)^2 h
=
\Bigl(\frac{D}{ds}\Bigr)^2 h
=
\frac{D^2 H}{ds^2} \cdot H^{-1}\cdot h
+ 2 \frac{D H}{ds} \cdot
\int^{r_s^{\prime\prime}}_{r_s^\prime }
\bigl( \frac{D\Lambda(u)}{ds} \cdot \dot\gamma_s(u)
+ \Lambda (u)\cdot \frac{D \dot\gamma_s(u)}{ds} \Bigr) du+
\]
\[
+H\cdot \int^{r_s^{\prime\prime}}_{r_s^\prime }
\Bigl( \frac{D^2\Lambda(u)}{ds^2} \cdot  \dot\gamma_{s}(u)
+2 \frac{D\Lambda(u)}{ds}  \cdot \frac{D\dot\gamma_s(u)}{ds}
+\Lambda (u)\cdot \frac{D^2\dot\gamma_s(u)}{ds^2}
\Bigr) du +\rho ,
\qquad (3.23)
\]
where
\[
\rho
:= H\cdot
\Big\{
\frac{D}{ds}\Big[ \frac{d r_s^{\prime\prime}}{ds}
\Lambda(r_s^{\prime\prime})\cdot \dot\gamma_{s}(r_s^{\prime\prime})
-  \frac{d r_s^{\prime}}{ds} \dot\gamma_{s}(r_s^{\prime})
\Big]
+
\frac{d r_s^{\prime\prime}}{ds}
\Bigl[
\frac{D}{ds} (\Lambda(u)\cdot \dot\gamma_{s}(u)
\Bigr]\Big|_{u=r^{\prime\prime}_{s}}-
\]
\[
- \frac{d r'_s}{ds}
\Bigl[\frac{D}{ds} (\Lambda (u)\cdot \gamma _{s}(u))\Bigr]
\Big|_{u=r^\prime_{s}}
\Big\}
+ 2 \frac{D H}{ds} \cdot
\Bigl[ \frac{d r_s^{\prime\prime}}{ds} \Lambda (r_s^{\prime\prime})
\cdot \dot\gamma_{s}(r_s^{\prime\prime})
-  \frac{d r'_s}{ds} \dot\gamma_{s}(r_s')
\Bigr]
\]
arises from the differentiation with respect to $s$ of the boundaries of
integration $r$ and $r$. Let us note that usually $[13-16]$ the statement
of the problem is such that $r$   and $r$   do not depend on $s$, therefore
$\rho =0$.

By its essence the equation (3.23) gives an answer to the problem stated
above. In particular, if we write the term $D^{2}H/ds^{2}$ in detail, we shall
see the "major" dependence of the relative acceleration $D^{2}h/ds^{2}$ on the
force $F$ acting on the observer. But more essential is the dependence on the
force field $F_{s}$a nd to write it we shall transform the derivative $D^{2}$
$/ds^{2}$ in (3.23) as follows.

Taking into account the evident equality $\nabla _{\gamma ^\prime }$  $=D$
$/ds$, from the basic equation (3.5) for $U=\gamma ^\prime $ and $\xi
=\dot\gamma$, we get
\[
D^{2}\dot\gamma/ds^{2}=\nabla ^{2}_{\gamma ^\prime }\dot\gamma
=R(\gamma^\prime, \dot\gamma) \gamma ^\prime
+\nabla _{\dot\gamma }(\nabla_{\gamma ^\prime }\gamma ^\prime )
+\nabla_{\gamma^\prime} (T(\gamma ^\prime,\dot\gamma))
+\nabla_{\gamma^\prime }[\gamma^\prime,\dot\gamma]
+\nabla _{[\gamma ^\prime,\dot\gamma] }\gamma ^\prime .
\]

	In this equality the last two terms are zeros because of
$[\gamma^\prime,\dot\gamma]=0$. (In fact, the i-th component of this
commutator at $\gamma _{s}(r)$ is
\[
([\gamma ^\prime, \dot\gamma]\mid _{\gamma _{s}(r)})^{i}
=(\gamma ^\prime (\dot\gamma^{i}_{s})
- \dot\gamma(\gamma^{\prime {i}} ))\mid _{\gamma _{s}(r)}
=\partial^{2}\gamma ^{i}_{s}(r)/\partial r\partial s-\partial ^{2}\gamma
^{i}_{s}(r)/\partial s\partial r\equiv 0,
\]
where we suppose a $C^{2}$dependence of $\gamma _{s}(r)$ on $s$ and r.) So,
using this, $(\nabla _{\gamma ^\prime }\gamma ^\prime )=D\gamma ^\prime
/ds\mid _{\gamma _{s}}=F_{s}(r)$ (see (3.20a)) and
$\nabla _{\dot\gamma }F_{s}(r)= =DF_{s}(r)/dr$, we find
\[
\frac{D^2\dot\gamma}{ds^2}\Big|_{\gamma_{s}(r)}
=R(\gamma ^\prime ,\dot\gamma)\gamma^\prime\mid _{\gamma_{s}(r)}
+\frac{d}{dr}F_{s}(r)
+T(F_{s},\dot\gamma_{s})\mid _{\gamma _{s}(r)}+
\]
\[
\Bigl(\frac{D T}{ds} (\gamma',\dot\gamma)\Bigr)
\Big|_{\gamma _{s}(r)}
+T(\gamma ^\prime ,\frac{D\dot\gamma}{ds})|_{\gamma_s(r)}
\]
and consequently (3.23) takes the form
\[
\Bigl( \frac{D}{ds}\Big|_x\Bigr)^2 h
=\frac{D^2H}{ds^2} \cdot H^{-1}\cdot h
+ 2\frac{DH}{ds} \cdot
\int^{r^{\prime\prime}_{s}}_{r_s^\prime }
\Bigl(
\frac{D\Lambda(u)}{ds}\cdot \dot\gamma_{s}(u)
+ \Lambda (u)\cdot \frac{D\dot\gamma_s(u)}{ds}
\Bigr)du +
\]
\[
+ H\cdot \int^{r_s^{\prime\prime}}_{r_s^\prime }
\Big\{
\frac{D^2\Lambda(u)}{ds^2} \cdot \dot\gamma_{s}(u)
+2\frac{D\Lambda(u)}{ds} \cdot \frac{D\dot\gamma(u)}{ds}
+\Lambda (u)\cdot
\bigl[\bigl(R(\gamma ^\prime,\dot\gamma)\gamma ^\prime
+\frac{D T}{ds}(\gamma ^\prime,\dot\gamma  ) +
\]
\[
+T(\gamma ^\prime, \frac{D\dot\gamma}{ds})\bigr)|_{\gamma_{s}(u)}
+\frac{D}{du}F_{s}(u)
+T(F_{s}, \dot\gamma_{s})|_{\gamma _{s}(u)}\bigr]du
+ \rho .
\qquad (3.24)
\]
This equation is the answer of the problem stated in this example problem.
It represents the deviation equation in the form of an equation of motion in
the considered case.

From a dynamical point of view the most important terms in (3.24) are those
containing explicitly the force $F_{s}(r)$, i.e.
\[
H\cdot \int^{r_s^{\prime\prime}}_{r_s^\prime }\Lambda (u)\cdot
\bigl[ \frac{D}{du} F_{s}(u)
+T(F_{s}, \dot\gamma_{s})|_{\gamma_{s}(u)}\bigr]du
=L^{\eta_{s}}_{t^\prime_{s}  \to t^{\prime\prime}_{s}}
(L^{\gamma_{s}}_{r^{\prime\prime}_{s}  \to r^\prime _{s}}
F_{s}(r_s^{\prime\prime}  ) -
\]
\[
-F_{s}(r'_s))
+ H\cdot \int^{r+s^{\prime\prime}}_{r_s^\prime }[\Lambda(u)
\cdot T(F_{s},\dot\gamma_{s})|_{\gamma _{s}(u)}
-  \frac{D\Lambda(u)}{ds} \cdot F_{s}(u)]du,
\qquad (3.25)
\]
where we have done an evident integration by parts of the integrand
$\Lambda (u)\cdot\frac{D}{du}F_{s}(u)$. Let us note that the first term in
(3.25), which is written explicitly by a transport $L$ is simply the
difference defined by means of $L$ at the point $x(s)$ of the forces
$F_{s}(r^{\prime\prime})$ and $F_{s}(r^\prime )$ acting on the observed
particles.

At the end, we are going to consider two important special cases of (3.24).

First, in the Euclidean case (3.24) reduces to the second law of the
Newtonian mechanics. In fact, in this case we can put $M={\Bbb R}^{n}$,
$dr'_s/ds=dr_s^{\prime\prime}/ds=0$ and $H=\Lambda (u)=\delta $, where
$\delta $ is the unit tensor with components the Kroneker deltas
$\delta^{i}_{j}$ (see $(1.2^\prime ))$, and if we use a basis in which
$\Gamma ^{i}_{.jk}=0$, then (3.24) becomes
\[
\frac{d^2h}{ds^2}
=
\Bigl(\frac{D^2}{ds^2}\Big|_x\Bigr) h
= \int^{r_s^{\prime\prime}}_{r_s^\prime } \frac{d}{du} F_{s}(u)du
=F_{s}(r_s^{\prime\prime})-F_{s}(r_s^\prime ).
\qquad (3.26)
\]
Second, in the infinitesimal case (3.24) reduces to the equation known, e.g.
from [16], for the relative motion of two "sufficiently near" point
particles.

For brevity and simplicity we shall suppose
$dr'_s/ds=dr_s^{\prime\prime}/ds=0$. As a consequence of (1.2), we have
\[
H=\delta +O(t_s^{\prime\prime}  - t_s^{\prime}),
\quad H^{-1} =\delta + O(t_s^{\prime\prime}  - t_s^{\prime}),
\quad \Lambda (r)=\delta +O(r-r'_s).
\]

	Using these equalities, the formula
\(
\int^{r^{\prime\prime}}_{r^\prime }f(u)du
=f(r^\prime )(r^{\prime\prime}-r^\prime )
+
O((r^{\prime\prime}-r^\prime )^{2})
\)
for any $C^{1}$ function $f:[r^\prime ,r^{\prime\prime}]  \to {\Bbb R}$, and
the infinitesimal deviation vector
$\zeta :=\zeta (s):=\dot\gamma_{s}(r'_s )(r_s^{\prime\prime} -r_s^{\prime})$
(see~(2.10)) from (3.24), we obtain
\[
\frac{D^2\zeta}{ds^2}
=R(\gamma ^\prime ,\zeta )\gamma^{\prime}|_{\gamma _{s}(r')}
+ \frac{DF_s(r)}{dr}\big|_{r=r^\prime }
(r^{\prime\prime}_s - r^{\prime}_s)
+ T(F_{s},\zeta )|_{\gamma(r'_{s})}+
\]
\[
+ \frac{DT}{d s} (\gamma ^\prime ,\zeta )\Big|_{\gamma_{s}(r'_s)}
+ T(\gamma ^\prime ,  \frac{D\zeta}{d s} )|_{\gamma _{s}(r'_s)}
+ O(t_s^{\prime\prime}-t_s^\prime ) +
O((r_s^{\prime\prime}-r_s^\prime )^{2}).
\qquad(3.27)
\]

If here we neglect the terms $O(t_s^{\prime\prime}-t_s^\prime)$ and
$O((r_s^{\prime\prime}-r_s^\prime)^{2})$ and put $T=0$, we get the equation
derived in [16], ch.~8, sect.~1 for relative motion of two "nearly" moving
point particles.

\medskip
\medskip
\noindent {\bf 4. CONCLUDING REMARKS}

\medskip
The displacement vector introduced in Sect. 2 is a direct generalization of
the difference of two Euclidean (radius-)vectors. To show this, we consider
the (pseudo-)Euclidean transport generated by Cartesian coordinates in
$M={\Bbb R}^{n}$ or $M=E^{n}$, which is insignificant now (see [2], definition
3.1), i.e. as a concrete realization of I we shall use the parallel transport
in ${\Bbb R}^{n}$ will be used. Then in any basis, we have
$(I^{\gamma }_{s\to t}u)^{i}=u^{i}$ for any path
$\gamma :J  \to {\Bbb R}^{n}$, every $u\in T_{\gamma (s)}({\Bbb R}^{n})$ and
arbitrary $s,t\in $J. Hence in this case (1.1b) gives
\[
(d^{\gamma }_{s}(t))^{i}
:= \int_{s}^{t} \dot\gamma ^{i}(u)du
=\gamma^{i}(t)-\gamma ^{i}(s),
\quad  s,t\in J,\qquad (4.1)
\]
which proves the above statement.

As it is known to the author, the equality (2.5) is published for the first
time in [7] (see therein equation (1) in which a slightly different notation
is used). Its full derivation in local coordinates, with the usage of Lie
derivatives, is presented in [8] (see therein section 1 and the appendix).
More precisely, in [7] the equation (2.8) is given for an arbitrary path $x$
and vector $h_{21}$ (with the usage of $[U,h_{21}]=L_{U}h_{21}$ and
$F:=\nabla _{U}U)$, the proof of which has been published later in [8]. As a
consequence of the arbitrariness of $x$ and $h_{21}$, for which in
$[8] (2.8)$ is proved, in this case the qualities (2.8) and (2.5) are
equivalent.

Independently the equality (2.5) is found in [14] from where is taken the
presented here its derivation.

\medskip
\medskip
\noindent {\bf ACKNOWLEDGEMENT}

\medskip
This research was partially supported by the Found for Scientific Research of Bulgaria under contract Grant No. $F 103$.

\medskip
\medskip
\noindent {\bf REFERENCES}

\medskip
1.  Iliev B.Z., Transports along paths in fibre bundles. General theory, Communication JINR, $E5-93-299$, Dubna, 1993.\par
2.  Iliev B.Z., Linear transports along paths in vector bundles. I. General theory, Communication JINR, $E5-93-239$, Dubna, 1993.\par
3.  Kobayashi S., K. Nomizu, Foundations of differential geometry, vol.1, Interscience publishers, New-York-London, 1963.\par
4.  Dubrovin B.A., S.P. Novikov, A.T. Fomenko, Modern geometry, Nauka, Moscow, 1979 (In Russian).\par
5.  Iliev B.Z., Deviation equations in spaces with affine connection. I. Generalized deviation equation: physical interpretation, local case and nonlocal problems. II. Displacement vector: the general case, Bulgarian Journal of Physics, vol.13, No.$6, 1986, pp.494-506$.\par
6.  Iliev B.Z., Generalized transports and displacement vectors, Communication JINR, $E2-87-267$, Dubna, 1987.\par
7.  Iliev B.Z., S. Manov, Deviation equations in spaces with torsion, In: Proceedings of the 5-th Soviet (USSR) Gravity Conference "Modern theoretical and experimental problems of relativity theory and gravitation", Moscow Univ., Moscow, 1981, p.122 (In Russian).\par
8.  Iliev B.Z., S. Manov, Deviation equations in spaces with affine connection, Communication JINR, $P2-83-897$, Dubna, 1983 (In Russian).\par
9.  Iliev B.Z., Deviation equations in spaces with affine connection. III. Displacement vector: the case of a space with affine
connection. IV. Nonlocal deviation equation, Bulgarian Journal of Physics, vol.14, No.$1, 1987, pp.18-31$.\par
10.  Iliev B.Z., The deviation equation as an equation of motion, Communication JINR, $E2-87-78$, Dubna, 1987.\par
11.  Iliev B.Z., Linear transports along paths in vector bundles. II. Some applications, Communication JINR, $E5-93-260$, Dubna, 1993.\par
12.  Iliev B. Z., Special bases for derivations of tensor algebras. II. Case along paths, Communication JINR, $E5-92-508$, Dubna, 1992.\par
13.  Synge J.L., Relativity: The general theory, North-Holland Publ. Co., Amsterdam, 1960.\par
14.  Swaminarayan N.S., J.L. Safko, A coordinate-free derivation of a generalized geodesic deviation equation, J. Math. Phys., {\bf 24}, No.$4, 1983, pp.883-885$.\par
\noindent 15.  Manoff S., Gen. Relativ. Gravit., ${\bf 1}{\bf 1}{\bf ,} 1979, pp.189-204$. \par
16.  Weber J., General Relativity and Gravitational Waves, New York, 1961.

\newpage

\medskip
\medskip
\noindent Iliev B. Z.\\[5ex]

Deviation Equations in Spaces with a Transport along Paths\\[5ex]

\medskip
The displacement and deviation vectors in spaces (manifolds), the tangent
bundle of which is endowed with a transport along paths, are introduced. In
case these spaces are equipped with a linear connection, the deviation
equations (between arbitrary, geodesic or not, paths) in such spaces are
investigated.\\[5ex]

\medskip
\medskip
The investigation has been performed at the Laboratory of Theoretical
Physics, JINR.

\end{document}